# Ideal-Gas Approach to Hydrodynamics

Zhe-Yu Shi[1,*] Chao Gao[2,†] and Hui Zhai[3,‡]

[1]*State Key Laboratory of Precision Spectroscopy, East China Normal University, Shanghai 200062, China*
[2]*Department of Physics, Zhejiang Normal University, Jinhua 321004, China*
[3]*Institute for Advanced Study, Tsinghua University, Beijing 100084, China*

Transport is one of the most important physical processes in all energy and length scales. Ideal gases and hydrodynamics are, respectively, two opposite limits of transport. Here, we present an unexpected mathematical connection between these two limits; that is, there exist situations that the solution to a class of interacting hydrodynamic equations with certain initial conditions can be exactly constructed from the dynamics of noninteracting ideal gases. We analytically provide three such examples. The first two examples focus on scale-invariant systems, which generalize fermionization to the hydrodynamics of strongly interacting systems, and determine specific initial conditions for perfect density oscillations in a harmonic trap. The third example recovers the dark soliton solution in a one-dimensional Bose condensate. The results can explain a recent puzzling experimental observation in ultracold atomic gases by the Paris group and make further predictions for future experiments. We envision that extensive examples of such an ideal-gas approach to hydrodynamics can be found by systematical numerical search, which can find broad applications in different problems in various subfields of physics.

## I. INTRODUCTION

Studying transport of matter is an important subject in almost all subfields of physics, ranging from structure formation in astrophysics on cosmological scales [1] to collective motions of electrons in solid-state materials on microscopic scales [2], from quark-gluon plasma as the highest-temperature quantum matters created in colliders [3] to ultracold atomic gases realized at the lowest temperature in laboratories [4]. It is well known that, for these transport phenomena in different systems, there are two opposite limits known as the hydrodynamic regime and the collisionless regime.

The distinction between these two regimes crucially depends on the relaxation time $\tau_r$ of the system [5]. If the relaxation time $\tau_r$ is much shorter than the typical dynamical timescale $\tau_d$, the transport is said to be in the hydrodynamic regime. In this regime, the system can retain local equilibrium during the dynamical process and, therefore, is well described by its local density distribution $n(\mathbf{r}, t)$ and local velocity distribution $\mathbf{v}(\mathbf{r}, t)$ governed by a set of hydrodynamic equations. In the opposite limit, when $\tau_r$ is much longer than $\tau_d$, the system is said to be in the collisionless regime. In this regime, the system cannot reach local equilibrium during the dynamical process, and, therefore, descriptions of the system have to involve the Wigner function $f(\mathbf{r}, \mathbf{p}, t)$, which describes how the momentum distribution at position $\mathbf{r}$ follows time evolution. Ideal gas is one such example whose dynamics is described by $f(\mathbf{r}, \mathbf{p}, t)$. Once $f(\mathbf{r}, \mathbf{p}, t)$ is known, the time dependence of $n(\mathbf{r}, t)$ and $\mathbf{v}(\mathbf{r}, t)$ can be deduced from the Wigner function. It is well-accepted conventional wisdom that the physics is drastically different between these two regimes.

In this work, we report a result that is sharply in contrast to the conventional wisdom. We find that the solutions to a class of hydrodynamic equations with certain initial conditions can be directly constructed from the solutions to the dynamics of ideal gases. The main finding is that, for certain initial conditions, when the one-particle Liouville equation for ideal gases is formally recast into the form of hydrodynamic equation, the "formal pressure tensor" depends only on local density $n(\mathbf{r}, t)$ and does possess the physical meaning as the real pressure of another interacting system. This unexpected connection between these two opposite limits of transport is schematically illustrated in Fig. 1.

Below, we first state rigorously the content of the ideal-gas approach to hydrodynamics and then discuss various physical realizations of such an approach, as summarized in Table I. The first two examples focus on scale-invariant systems. The first example can be viewed as a generalization of the fermionization in a strongly interacting one-dimension system [6,7] to the hydrodynamics level. The second example finds out a geometric description of the specific initial conditions in two and three dimensions, under which the real space density distribution undergoes perfect breathing oscillation in a harmonic trap. This provides a physical explanation for a puzzling experimental discovery in ultracold atomic gases reported by the Paris group [8]. In this experiment, a unique perfect oscillation of density distribution, called a *breather*, is observed for equilateral triangular initial density distribution but not for other polygonal initial densities [8]. This phenomenon is also confirmed by numerical simulation of the Gross-Pitaevskii equation [8,9], but it still lacks a physical understanding why the equilateral triangular shape is special, which is now answered by our geometric condition. Finally, the third example recovers the well-known dark soliton solution in one-dimensional superfluids [10,11], indicating that the connections can be found in a more broad context beyond scale-invariant systems.

*zyshi@lps.ecnu.edu.cn
†gaochao@zjnu.edu.cn
‡hzhai@tsinghua.edu.cn

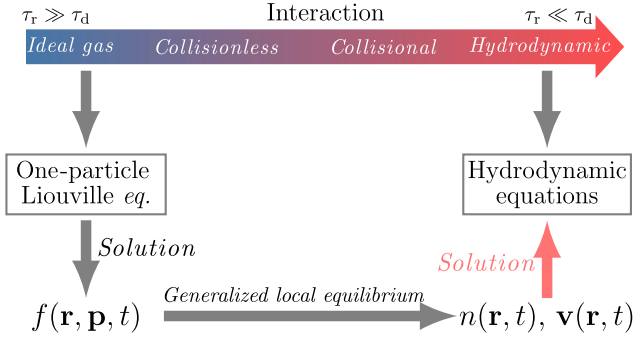

FIG. 1. Ideal-gas approach hydrodynamics. The one-particle Liouville equation for ideal gases and the hydrodynamic equations for interacting systems, respectively, work for two opposite limits. Here, we show, unexpectedly, that the solution to certain one-particle Liouville equation of ideal gases can be used to construct the solution to a set of hydrodynamic equations if a generalized local-equilibrium condition [Eq. (8)] can be satisfied.

## II. THE FORMALISM

We start with the Boltzmann equation in $D$ dimension:

$$\partial_t f + \frac{1}{m}\mathbf{p} \cdot \nabla_\mathbf{r} f - (\nabla_\mathbf{r} U) \cdot \nabla_\mathbf{p} f = I[f], \quad (1)$$

where $m$ is the particle mass, $U(\mathbf{r})$ is an external potential, and $I[f]$ represents the collision term. By solving Eq. (1), we can obtain the Wigner function $f(\mathbf{r}, \mathbf{p}, t)$, with which we can construct the density and velocity distributions as

$$n(\mathbf{r}, t) = \int d^D\mathbf{p} f(\mathbf{r}, \mathbf{p}, t), \quad (2)$$

$$\mathbf{v}(\mathbf{r}, t) = \frac{\langle \mathbf{p} \rangle}{mn(\mathbf{r}, t)} = \frac{1}{mn(\mathbf{r}, t)} \int d^D\mathbf{p} \mathbf{p} f(\mathbf{r}, \mathbf{p}, t). \quad (3)$$

Here, we introduce $\langle \ldots \rangle$ as the local density of a quantity that is averaged under the Wigner function:

$$\langle O(\mathbf{p}) \rangle \equiv \int \frac{d^D\mathbf{p}}{(2\pi)^D} O(\mathbf{p}) f(\mathbf{r}, \mathbf{p}, t), \quad (4)$$

where $O(\mathbf{p})$ is a function of momentum. We can formally introduce a tensor $P_{\alpha\beta}$ as

$$P_{\alpha\beta} = \frac{1}{m}\left(\langle p_\alpha p_\beta \rangle - \frac{1}{n}\langle p_\alpha \rangle \langle p_\beta \rangle\right), \quad (5)$$

where the subindices $\alpha, \beta = 1, 2, \ldots, D$ stand for the $\alpha$th or $\beta$th component of the corresponding vector. If tensor $P_{\alpha\beta}$ is proportional to the identity matrix, i.e., $P_{\alpha\beta} = P\delta_{\alpha\beta}$, then it can be shown that the density and velocity fields constructed by Eqs. (2) and (3) satisfy the following two equations [12,13]:

$$\partial_t n + \nabla \cdot (n\mathbf{v}) = 0, \quad (6)$$

$$m(\partial_t \mathbf{v} + \mathbf{v} \cdot \nabla \mathbf{v}) + \nabla U + \frac{1}{n}\nabla P = 0, \quad (7)$$

which are essentially the particle number and the momentum conservation, respectively. (See the Appendix A for the derivation.)

TABLE I. Summary of examples of physical realizations of the *ideal-gas approach to hydrodynamics* discussed in this paper. As illustrated by the table, these examples include different systems in different dimensions, and the diverse phenomena in these interacting systems can all be obtained by solutions of the Liouville equation. In the pressure column, $P[n]$ denotes how the pressure depends on the density $n$, and $g$ is the interaction parameter for different systems.

|  | Dimensionality | Pressure | Phenomenon |
|---|---|---|---|
| Example I | 1D | $P(n) = (g/m)n^3$ | Generalized fermionization |
| Example II | 2D/3D | $P(n) = (g/m)n^2/(g/m)n^{5/3}$ | 2D/3D perfect breather |
| Example III | 1D | $P[n] = \frac{1}{2}gn^2 - (n/4m)\partial_x^2 \log(n)$ | Dark soliton |

Generally, $P_{\alpha\beta}$ defined by Eq. (5) is a function of $\mathbf{r}$ and $t$ [13]. However, in order for Eq. (7) to represent the Euler equation in hydrodynamics, it is crucial that $P_{\alpha\beta}$ possesses the meaning of local pressure, usually being a functional of $n(\mathbf{r}, t)$ not only for all spatial points but also at all time, that is,

$$P_{\alpha\beta} = P[n]\delta_{\alpha\beta}. \quad (8)$$

The standard approach usually assumes that the relaxation time $\tau_r$ is short enough such that the Wigner function $f(\mathbf{r}, \mathbf{p}, t)$ is not far away from local equilibrium, and, therefore, it ensures Eq. (8) being the equilibrium pressure of the system.

In the following, we present a quite different situation to satisfy Eq. (8) without this local-equilibrium assumption. The resulting $P_{\alpha\beta}$ can still acquire the meaning of the equilibrium pressure but of another different system. In this sense, we call Eq. (8) a *generalized local-equilibrium* (GLE) condition. We show that the GLE condition can be satisfied by a noninteracting ideal gas, where the relaxation time is infinity. When $I[f]$ in Eq. (1) vanishes for a noninteracting ideal gas, Eq. (1) is nothing but the one-particle Liouville equation (denoted as the Liouville equation in this paper below). It is worth emphasizing that, in these cases, the Liouville equation and the hydrodynamic equations describe two drastically different systems. The Liouville equation describes the dynamics of an ideal gas whose solution can be obtained exactly, and the hydrodynamic equations describe an interacting system that retains local equilibrium. However, once the Wigner function of the ideal gas satisfies the GLE condition, the solution of the corresponding hydrodynamic equations can then be obtained exactly from the dynamics of ideal gases, which is now called the *ideal-gas approach to hydrodynamics*.

We note that there is a physical intuition why the Boltzmann equation can be related to dissipationless hydrodynamic equations. According to H theorem, the Boltzmann equation breaks the time-reversal symmetry, and the system entropy keeps increasing over time when the collision integral $I[f]$ is finite. However, the dissipationless hydrodynamics always respects the time-reversal symmetry. Thus, these two equations are compatible only in two cases, that are (i) when the interaction is strong enough to ensure local equilibrium and (ii) when the system is noninteracting and the collision term is absent. The standard textbook approach considers the first case, and here we employ the second one.

As we discuss in the introduction, the Liouville equation lies on the opposite limit to the hydrodynamic equations; thus, a natural concern is whether the GLE condition can be satisfied by an ideal gas without local equilibrium and whether the resulting pressure function describes an actual physical system. Below, we present a number of examples, which provide positive answers to resolve these concerns.

These examples are summarized in Table I. As one can see, these examples reveal intriguing phenomena related to experimental observations.

## III. EXAMPLE I

In the first example, we consider ideal gas in a one-dimensional harmonic trap $U(x) = m\omega^2 x^2/2$, and we consider an initial Wigner function given by ($\hbar = 1$)

$$f_0(x, p) = \frac{n_0(x)}{2k_F}\Theta(k_F - |p|), \quad (9)$$

with $n_0(x) = \rho_0\Theta(L/2 - |x|)$. Here, $\Theta(x)$ stands for a unit step function, and $k_F$, $L$, and $\rho_0 > 0$ are free parameters. This initial Wigner function represents a uniform distribution inside the dashed rectangle in the $(x, p)$ phase space as plotted in Fig. 2(a), and the density $n_0(x)$ is uniformly distributed inside interval $[-L/2, L/2]$. It is worth mentioning that the present choice of $n_0(x)$ is just for simplicity; one can replace it with any convex function with $\partial_x^2 n_0(x) \leq 0$ and the following results still hold (see Appendix B for details).

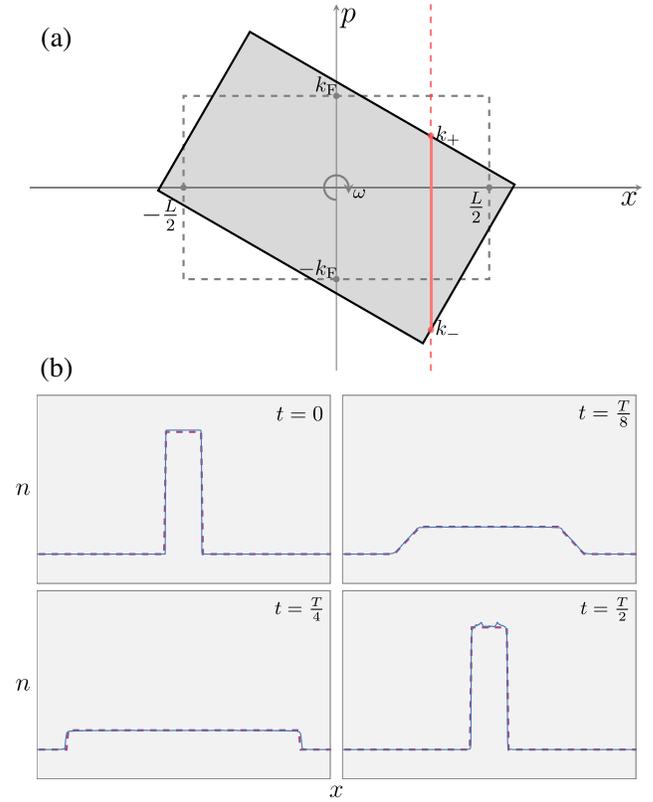

FIG. 2. Example I. (a) Schematic plot of the initial Wigner function, which is uniformly distributed inside the dashed rectangle, and the Wigner function at a finite time, which is uniformly distributed inside the solid rectangle. These two rectangles are related by a rotation with angle $\omega t$. The red line and $k_\pm$ are used in Appendix B for calculating momentum distribution at a given location $x$. (b) Density profiles $n(x, t)$ at $t = 0, T/8, T/4, T/2$ obtained by numerically solving the generalized Gross-Pitaevskii equations (blue solid lines) and the exact solution constructed from the ideal-gas dynamics (red dashed lines).

In the presence of a harmonic trap, the solution to the Liouville equation is given by $f(x, p, t) = f_0[x(t), p(t)]$, with

$$x(t) = x\cos(\omega t) - \frac{p}{m\omega}\sin(\omega t), \quad (10)$$

$$p(t) = p\cos(\omega t) + m\omega x \sin(\omega t). \quad (11)$$

As illustrated in Fig. 2(a), this corresponds to a rotation of the Wigner function by angle $\omega t$ in the phase space $(x, p)$. At finite time $t$, the Wigner function is then a uniform distribution inside the solid rectangle. It can be shown that this solution satisfies the GLE condition and gives rise to a pressure term $P(n) = gn^3/m$, with $g = k_F^2/(3\rho_0^2)$. (See Appendix B for the derivation.)

When the interaction energy of a quantum system scales in the same way as its kinetic energy, it leads to a scale-invariant quantum system whose equilibrium pressure at zero temperature takes a universal form:

$$P(n) = \frac{g}{m} n^{(D+2)/D}. \quad (12)$$

Thus, for $D = 1$, $P(n) = gn^3/m$ is the equilibrium pressure for a scale-invariant system [6,7,14,15]. This means that the hydrodynamic solution for a trapped one-dimensional scale-invariant system can be constructed by the Wigner function of free particles. Note that the initial Wigner function [Eq. (9)] resembles the momentum distribution of a one-dimensional Fermi sea. Thus, this result is reminiscent of the celebrated fermionization for the strongly interacting Tonk-Girardeau gas [6,7], although this one is on the hydrodynamic equation level and the other is on the microscopic many-body wave-function level.

We further confirm this result by showing the nice agreement between the exact solution constructed by the Wigner function and the numerical solution of a generalized Gross-Pitaevskii equation in Fig. 2(b). The details of the numerical solution are discussed in Appendix D. We note that the density distribution shows a perfect periodical behavior with a period $T/2$ with $T = 2\pi/\omega$. This is because, for the free evolution of particles in a harmonic trap, one always has $n(x, t + T/2) = n(-x, t)$. Together with the inversion symmetry, it naturally gives $n(x, t + T/2) = n(x, t)$. Here, we also note that the small discrepancy between the ideal-gas result and the numerical result is due to the so-called quantum pressure term, which exists in the generalized Gross-Pitaevskii equation but not in the hydrodynamics.

## IV. EXAMPLE II

The discussion of the first example can be generalized to higher dimensions. In two dimensions, we start with an initial Wigner function $f_0(\mathbf{r}, \mathbf{p})$ given by

$$f_0(\mathbf{r}, \mathbf{p}) = \frac{4\rho_0}{\sqrt{3}k_F^2} \Delta_L(\mathbf{r}) \nabla_{k_F}(\mathbf{p}). \quad (13)$$

Here, we use a simplified notation $\Delta_L(\mathbf{r})$ to represent a generalized two-dimensional $\Theta$ function, i.e., $\Delta_L(\mathbf{r}) = 1$ if $\mathbf{r}$ sits inside an equilateral triangle with side length $L$ and centered at the origin, and $\Delta_L(\mathbf{r}) = 0$ otherwise. $\nabla_{k_F}(\mathbf{p})$ is similar, which is unity when $\mathbf{p}$ sits inside an inverted equilateral triangle with side length $k_F$ and centered at the origin. This Wigner function gives rise to an initial density distribution that is uniform inside a triangle, i.e.,

$$n_0(\mathbf{r}) = \rho_0 \Delta_L(\mathbf{r}). \quad (14)$$

The solution to the Liouville equation also follows free evolution in a harmonic trap, which is given by

$$f(\mathbf{r}, \mathbf{p}, t) = f_0[\mathbf{r}(t), \mathbf{p}(t)], \quad (15)$$

with $\mathbf{r}(t) = \mathbf{r}\cos\omega t - (\mathbf{p}/m\omega)\sin\omega t$ and $\mathbf{p}(t) = \mathbf{p}\cos\omega t + m\omega\mathbf{r}\sin\omega t$. With $f_0$ given by Eq. (13), we can write

$$\begin{aligned} f(\mathbf{r}, \mathbf{p}, t) &= \frac{4\rho_0}{\sqrt{3}k_F^2} \Delta_L[\mathbf{r}(t)] \cdot \nabla_{k_F}[\mathbf{p}(t)] \\ &= \frac{4\rho_0}{\sqrt{3}k_F^2} \nabla_{m\omega L/\sin\omega t}(\mathbf{p} - m\omega\mathbf{r}\cot\omega t) \\ &\quad \times \nabla_{k_F/\cos\omega t}(\mathbf{p} + m\omega\mathbf{r}\tan\omega t) \\ &= \frac{4\rho_0}{\sqrt{3}k_F^2} \nabla_{k_s}(\mathbf{p} - \mathbf{p}_0), \end{aligned} \quad (16)$$

where we assume $t \in (0, T/4)$ such that both $\sin\omega t$ and $\cos\omega t$ are positive.

Crucially, the last equality in Eq. (16) follows the fact that the overlapped area of two homothetic equilateral triangles is still an equilateral triangle, whose center and side length are denoted by $\mathbf{p}_0$ and $k_s$, respectively. Both $\mathbf{p}_0$ and $k_s$ are functions of $\mathbf{r}$ and $t$. In fact, it can be shown that, independent of their sizes and relative positions, the overlap area of two homothetic equilateral triangles is always of the same shape, up to a scaling and translation. As shown in Fig. 3(a), this geometric property is unique for triangles and does not hold for other polygons in two dimensions [16]; for instance, the overlapped area of two squares or two hexagons is not a square or hexagon, which, in general, depends on the relative positions and sizes of these two. As a result of this unique geometric property, we see that the Wigner functions at different positions and time, up to scaling and translation, always follow the same distribution, i.e.,

$$f(\mathbf{r}, \mathbf{p}, t) = u[\lambda(\mathbf{p} - \mathbf{p}_0)], \quad (17)$$

with $u(\mathbf{p}) = (4\rho_0/\sqrt{3}k_F^2)\nabla_{k_F}$ and $\lambda = k_F/k_s$.

In Appendix C, we prove that, if up to a scaling and a translation, the Wigner function $f(\mathbf{r}, \mathbf{p}, t)$ at different position and time is a universal function $u(\mathbf{p})$, i.e., it satisfies Eq. (17), where both the scaling factor $\lambda$ and the momentum center $\mathbf{p}_0$ can be functions of $\mathbf{r}$ and $t$, and, moreover, if the second moment of $u(\mathbf{p})$, with respect to its center of mass, is proportional to an identity matrix, then $f(\mathbf{r}, \mathbf{p}, t)$ satisfies the GLE condition and it results in a pressure term as given by Eq. (12) for scale-invariant systems. We note that, in two dimensions, only the equilateral triangle has an isotropic second moment, even though all triangles share the geometric property that the intersection of two homothetic copies is homothetic to itself.

Thus, following the dynamics of the ideal gas, we can obtain the exact solution to the hydrodynamic equations of a two-dimensional scale-invariant system within $t \in (0, T/4)$, if the initial density is uniformly distributed inside an equilateral triangle. To determine the dynamics beyond $T/4$, we note that the exact solution of Eq. (16) shows that, at $T/4$, the density is uniformly distributed inside an inverted equilateral triangle, and the velocity field is zero everywhere. Thus, another similar Wigner function can be constructed for solving hydrodynamic equations in the time interval $t \in (T/4, T/2)$. Considering free evolution in a harmonic trap, the motion during $t \in (T/4, T/2)$ can be viewed as an inverted process for $t \in (0, T/4)$. In fact, it is not difficult to show $n(\mathbf{r}, T/4 + t) = n(\mathbf{r}, T/4 - t)$ as a result of the time-reversal symmetry. Hence, the density and velocity distributions recover their initial distributions at $t = T/2$. Repeating this construction, we thus find a periodical solution to the hydrodynamic equations with a period of $T/2$.

In Fig. 3(b), we again show nice agreements between the exact solution constructed from the ideal-gas dynamics and the numerical solution to the two-dimensional Gross-Pitaevskii equations. We also show that, for other initial polygonal geometries such as a square and a hexagon, numerical simulations do not find the periodical oscillation of the density distribution, consistent with the fact that Wigner functions evolving from other initial polygonal states do not satisfy the GLE condition. This result explains a recent experiment in a two-dimensional scale-invariant Bose condensate by the Paris group, where a unique perfect oscillation of density distribution, called a breather, is observed for equilateral triangular initial density distribution but not for other polygonal initial densities [8,17].

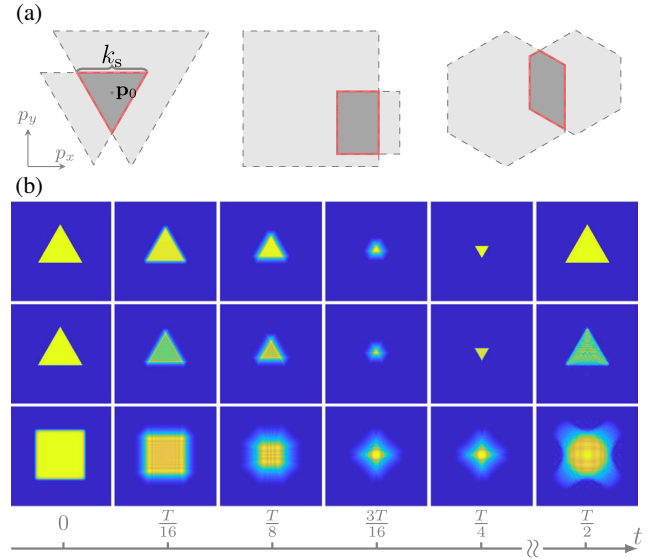

FIG. 3. Example II in 2D. (a) Schematic plots showing the unique property of triangles. The overlapped area of two homothetic equilateral triangles, as enclosed by the red triangle, is always another equilateral triangle, despite the sizes and relative position of these two. This geometric property does not hold for other shapes such as two homothetic squares or hexagons. (b) Dynamics of density distributions with initial density uniformly distributed inside an equilateral triangle, obtained by exact construction from the ideal-gas dynamics (top row) or by numerically solving the Gross-Pitaevskii equations (middle row), and the same dynamics with initial density uniformly distributed inside a square (lower row).

This phenomenon is also confirmed by numerical simulation of the Gross-Pitaevskii equation [8,9].

These discussions can be straightforwardly generalized to three dimensions. In this case, we should first search for three-dimensional geometric objects that satisfy the same geometric property, i.e., the intersection of two homothetic objects is the same homothetic object. As shown in Fig. 4(a), the only possible geometric object is a tetrahedron. Following the same construction, we can then start with an initial Wigner function similar to Eq. (13), with the equilateral triangle replaced by a standard tetrahedron, as shown in Fig. 4(b). Similar to the derivation above, one can show that, in the presence of a harmonic trap, the time evolution of the Wigner function automatically obeys Eq. (17), and the resulting $u(\mathbf{p})$ has an isotropic second moment, which ensures that $f(\mathbf{r}, \mathbf{p}, t)$ satisfies the GLE and leads to a pressure of the three-dimensional scale-invariant system, such as the unitary Fermi gas [18]. Hence, the hydrodynamics of this three-dimensional scale-invariant system can also be exactly constructed from a Wigner function. This construction predicts that, at $t = T/4$, the density distribution is an inverted tetrahedron, which oscillates back to the initial tetrahedron at $t = T/2$. This prediction is confirmed by numerically

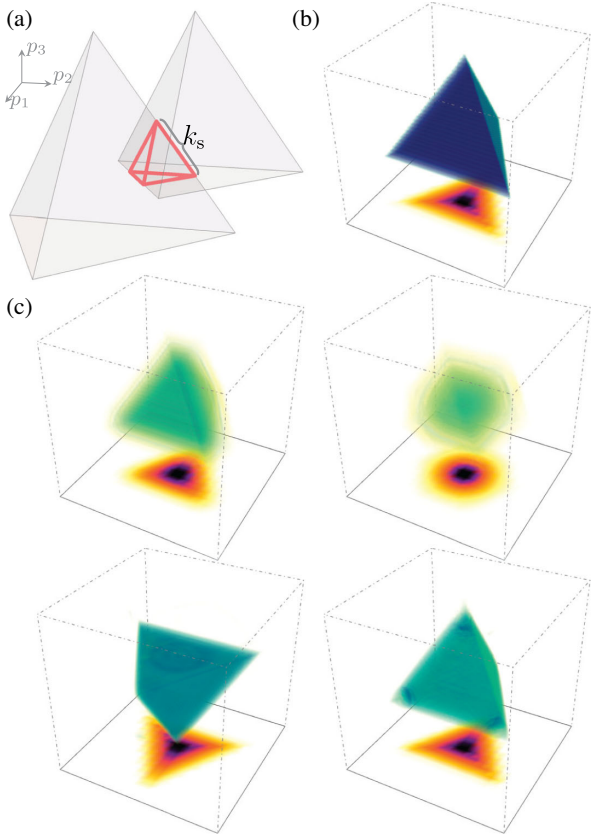

FIG. 4. Example II in 3D. (a) Schematic plots of the Wigner function uniformly distributed inside the intersection of two homothetic tetrahedra, and the intersection area, as enclosed by the red boundaries, is still a tetrahedron. (b) The initial density distribution. (c) Dynamics of the density distributions at $t = T/16, T/8, T/4, T/2$ obtained by numerically solving the generalized Gross-Pitaevskii equation with a scale-invariant pressure term in three dimensions.

solving the (generalized) Gross-Pitaevskii equation, as shown in Fig. 4(c).

## V. EXAMPLE III

The above two examples focus on scale-invariant systems. Here, we consider another example beyond systems with scale invariance. We consider a one-dimensional uniform system with $U(\mathbf{r}) = 0$ and a specific initial condition for the Liouville equation as

$$f_0(x, p) = n_0[\mathcal{N}_{m^2c^2/2}(p) - w(x)\delta(p - mu)], \quad (18)$$

with

$$w(x) = \frac{\sin^2 \gamma}{\cosh^2(x \sin \gamma/l)}. \quad (19)$$

Here, $\mathcal{N}_{m^2c^2/2}(p)$ is an arbitrary normalized distribution with zero mean and variance $m^2c^2/2$. $n_0$, $u$, and $c > 0$ are all free parameters, and we take $\cos \gamma = u/c$ and $l = 1/mc$.

For uniform systems, the solution to the Liouville equation is then given by $f(x, p, t) = f_0(x - pt/m, p)$, which leads to

$$f(x, p, t) = n_0[\mathcal{N}_{m^2c^2/2}(p) - w(x - ut)\delta(p - mu)]. \quad (20)$$

With $f(x, p, t)$ given by Eq. (20), straightforward calculations show that the GLE condition is satisfied, and the corresponding pressure functional is given by

$$P[n] = \frac{1}{2}gn^2 - \frac{n}{4m}\partial_x^2 \log n, \quad (21)$$

with $g = mc^2/n_0$. (See Appendix E for the derivation.) This specific pressure term corresponds to the one-dimensional superfluid hydrodynamic equations, which are ensured by long-range phase coherence and are equivalent to the one-dimensional Gross-Pitaevskii equation [4,19]

$$i\frac{\partial \psi}{\partial t} = -\frac{1}{2m}\partial_x^2 \psi + g|\psi|^2 \psi, \quad (22)$$

with $\psi = \sqrt{n}e^{i\varphi}$ and $v = \partial_x \varphi/m$. Thus, the initial density and velocity distributions correspond to a wave function $\psi(x)$ with a $2\gamma$ phase difference between $x = +\infty$ and $x = -\infty$ and a density dip at $x = 0$. Moreover, the density distribution at finite $t$ can be obtained straightforwardly from Eq. (20) as

$$n(x, t) = n_0\{\cos^2 \gamma + \sin^2 \gamma \tanh^2[(x - ut)\sin \gamma/l]\}. \quad (23)$$

It is, thus, clear that Eq. (23) is nothing but the well-known dark soliton solution of the one-dimensional Gross-Pitaevskii equation [10,11]. Hence, we arrive at an alternative picture for the dark soliton that can be mapped to dynamics of ideal gas, and the free evolution of noninteracting particles naturally explains why the density profile of a dark soliton can maintain its shape during its propagation.

## VI. OUTLOOK

In summary, this work establishes a surprising connection between the dynamics of free particles and the hydrodynamic equations of interacting systems. The key ingredient is that a generalized local-equilibrium condition can be satisfied by free evolution without physically reaching local equilibrium. A properly chosen initial condition for the free evolution plays a crucial role. Several examples are presented in this work, in which we analytically verify that the free particle dynamics can satisfy the GLE condition. For scale-invariant systems, we present a geometric condition for determining the specific condition, which also explains the recent puzzling experimental observations by the Paris group. We also show the dark-soliton example in one-dimensional superfluid, which means that our approach can be applied to general situations beyond scale-invariant systems. Regarding the generality of our approach, it contains the following two different aspects.

The first aspect is from the Liouville equation to the hydrodynamic equations. This can be studied by performing a systematic numerical search on the Liouville equation with various initial conditions, and then it is straightforward to verify whether the resulting tensor $P_{\alpha\beta}$ is a functional of density. If so, then the tensor $P_{\alpha\beta}$ can be interpreted as a pressure, and one can further ask whether there is a physical correspondence of this pressure term. It is conceivable that extensive numbers of cases can be found satisfying the GLE condition, and the corresponding hydrodynamic equations can, therefore, be solved exactly.

The other aspect is from the hydrodynamic equations to the Liouville equation, which is a more intriguing and difficult inverse problem, that is to say, given hydrodynamic equations with a fixed pressure term, whether we can find out a proper initial condition for the Liouville equation, which satisfies the GLE condition and yields precisely the same pressure. Our intuition based on constructing examples reported here is that this can be achieved for a large class of hydrodynamic equations. The reason is because, in hydrodynamic equations, the density and velocity field depend only on position $\mathbf{r}$, whereas in the Liouville equation, the Wigner function $f$ depends on both $\mathbf{r}$ and $\mathbf{p}$. There are multiple choices of $f(\mathbf{r}, \mathbf{p}, t)$ that can yield a given $n(\mathbf{r}, t)$ and $\mathbf{v}(\mathbf{r}, t)$, and it is conceivable that at least one such choice can satisfy the Liouville equation. Carefully designed numerical algorithms, perhaps with the help of machine learning, are needed to verify this conjecture. If this conjecture is correct, this can greatly simplify solving complex hydrodynamic equations and help to understand their behaviors. This potential developments can find broad applications in different branches of physics.

Finally, we note that a recent work establishes the relation between our solution of triangle breather and the Damski-Chandrasekhar shock wave [20]. Although this solution is exact to the hydrodynamic equation, we also note the difference between the hydrodynamic equation and the Gross-Pitaevskii equation. The latter is used in the numerical simulation and describes the real experiment. The quantum pressure term exists in the Gross-Pitaevskii equation but not in the hydrodynamic equation. It remains an open question that the initial evolution governed by the Gross-Pitaevskii equation chooses the Damski-Chandrasekhar shockwave to regularize the initial singularity [20].

## ACKNOWLEDGMENTS

We thank helpful discussions with Zhigang Wu, Jean Dalibard, Meera Parish, Jesper Levinsen, and Tianshu Deng. This work is supported by Program of Shanghai Sailing Program Grant No. 20YF1411600 (Z.-Y. S.), Beijing Outstanding Young Scientist Program (H. Z.), NSFC Grants No. 12004115 (Z.-Y. S.), No. 11835011 (C. G.), No. 12074342 (C. G.), and No. 11734010 (H. Z.), MOST under Grant No. 2016YFA0301600 (H. Z.), and Zhejiang NSFC Grant No. LY21A040004 (C. G.).

## APPENDIX A: DERIVING HYDRODYNAMIC EQUATIONS

In the following, we show that the density and velocity distributions given by Eqs. (2) and (3) satisfy both the continuity equation (6) and the Euler equation (7) under the GLE condition.

Note that the underlying collisions between microscopic particles conserve both the particle number and the center-of-mass momentum; therefore, it can be shown that the collision integral $I[f]$ in the rhs of the quantum Boltzmann equation must satisfy [12,13]

$$\int d^D\mathbf{p}\, I[f] = 0, \qquad \int d^D\mathbf{p}\, I[f]\mathbf{p} = 0. \quad (A1)$$

As a result, we can integrate the Boltzmann equation by momentum $\mathbf{p}$ and obtain

$$\int d^D\mathbf{p}\, \partial_t f(\mathbf{r}, \mathbf{p}, t) + \frac{1}{m}\nabla_\mathbf{r} \cdot \langle \mathbf{p} \rangle = 0. \quad (A2)$$

Together with the definitions in Eqs. (2) and (3), it immediately leads to the continuity equation (6). To verify the Euler equation, we can multiply the Boltzmann equation by $\mathbf{p}$ and integrate over both sides, which leads to

$$m\partial_t v_\alpha + m v_\beta \partial_\beta v_\alpha + \partial_\alpha U + \frac{1}{n}\partial_\beta P_{\alpha\beta} = 0, \quad (A3)$$

with $P_{\alpha\beta}$ defined by Eq. (5). When the GLE condition Eq. (5) is satisfied, i.e., when $P_{\alpha\beta} = P[n]\delta_{\alpha\beta}$, Eq. (A3) is then equivalent to the hydrodynamic Euler equation (7) with pressure $P[n]$.

## APPENDIX B: VERIFY THE GLE CONDITION FOR EXAMPLE I

As discussed in the main text and illustrated in Fig. 2(a), the time evolution of the Wigner function $f(x, p, t)$ corresponds to a rotation in the phase space $(x, p)$. Calculating the local momentum distribution $f(x, p, t)$ at a given position $x$ is then equivalent to calculating the intersection between the vertical line and the solid rectangle shown in Fig. 2(a). Assuming that the line intersects the rectangle at $k_+$ and $k_-$, $f(x, p, t)$ is then a uniform distribution between these two points, i.e.,

$$f(x, p, t) = \frac{\rho_0}{2k_F}\Theta(k_+ - p)\Theta(p - k_-). \quad (B1)$$

Straightforward calculation shows that

$$n = \rho_0 \frac{k_+ - k_-}{2k_F}, \qquad \langle p \rangle = n \frac{k_+ + k_-}{2}, \qquad (B2)$$

and the GLE condition is fulfilled because

$$\langle p^2 \rangle - \frac{1}{n}\langle p \rangle^2 = \frac{\rho_0(k_+ - k_-)^3}{24 k_F} = \frac{k_F^2}{3\rho_0^2} n^3 = g n^3, \qquad (B3)$$

where we use the fact that the second moment of a uniform distribution on the interval $[k_-, k_+]$ is $n(k_+ - k_-)^3/12$.

We note that the crucial point of the above proof is that the intersection between the red vertical line and the support of $f(x,p,t)$ is a singly connected interval $[k_-, k_+]$, such that the local momentum distribution at all $x$ and $t$ is a (shifted) Fermi distribution. This property remains true if we replace the rectangular support with other convex shapes in the phase space. For a system with a concave initial density distribution, i.e., $\partial_x^2 n_0 \leq 0$, we can write $f_0$ in a form similar to Eq. (9):

$$f_0(x,p) = \frac{n_0(x)}{2k_F(x)} \Theta[k_F(x) - |p|], \qquad (B4)$$

and here the key point is that a spatial-dependent Fermi momentum $k_F(x) = \sqrt{3g n_0(x)}$ is used to replace the constant $k_F$ in Eq. (9). It is clear that $f_0(x,p)$ is represented by a convex shape defined by $n_0(x)$ in the phase space. Thus, the previous derivation remains the same, and the GLE condition fulfills automatically.

## APPENDIX C: VERIFY THE GLE CONDITION FOR EXAMPLE II

Verifying the GLE condition in example II follows directly from the following proposition.

*Proposition.*—If, up to a scaling and a translation, the Wigner function $f(\mathbf{r},\mathbf{p},t)$ at different position and time is a universal function $u(\mathbf{p})$, i.e.,

$$f(\mathbf{r},\mathbf{p},t) = u[\lambda(\mathbf{p} - \mathbf{p}_0)], \qquad (C1)$$

where both the scaling factor $\lambda$ and the momentum center $\mathbf{p}_0$ can be functions of $\mathbf{r}$ and $t$, and if the second moment of $u(\mathbf{p})$, with respect to its center of mass, is proportional to an identity matrix, then $f(\mathbf{r},\mathbf{p},t)$ satisfies the GLE condition, and it results in a pressure term as given by Eq. (12) for scale-invariant systems.

To prove the proposition, we first assume the second moment of $u(\mathbf{p})$ is given by $u_2 \delta_{\alpha\beta}$, that is,

$$\langle p_\alpha p_\beta \rangle_u - \frac{1}{u_0}\langle p_\alpha \rangle_u \langle p_\beta \rangle_u = u_2 \delta_{\alpha\beta}, \qquad (C2)$$

where $\langle \ldots \rangle_u$ stands for momentum average over $u(\mathbf{p})$ and $u_0 = \int d^D \mathbf{p} u(\mathbf{p})$ is the zeroth moment of $u(\mathbf{p})$. Both $u_0$ and $u_2$ are constants. Then, both the density $n$ and the second moment of $f(\mathbf{r},\mathbf{p},t)$ can be calculated via a scaling transformation as

$$n = \int d^D \mathbf{p} u[\lambda(\mathbf{p} - \mathbf{p}_0)] = \lambda^{-D} u_0, \qquad (C3)$$

$$\langle p_\alpha p_\beta \rangle - \frac{1}{n}\langle p_\alpha \rangle \langle p_\beta \rangle = \lambda^{-(D+2)} u_2 \delta_{\alpha\beta} = g n^{(D+2)/D} \delta_{\alpha\beta}, \qquad (C4)$$

with $g = u_2/u_0^{(D+2)/D}$ a dimensionless constant independent of position $\mathbf{r}$ or time $t$. This completes the proof of this proposition.

The proposition indicates that the Wigner functions for cases satisfying the geometric condition discussed in example II also satisfy the GLE condition, because, in these cases, the Wigner functions can be expressed as Eq. (17). Moreover, this proposition requires that the second moment of $u(\mathbf{p})$ is proportional to an identity matrix. In two dimensions, even though all the triangles share the geometric property that the intersection of two homothetic copies is homothetic to itself, only the equilateral triangle can result in a function $u(\mathbf{p})$ that has an isotropic second moment. As a result, the only possible triangular breather mode is the equilateral one. In this case, $u(\mathbf{p}) = (4\rho_0/\sqrt{3}k_F^2)\nabla_{k_s}$, which gives $u_0 = \rho_0 k_s^2/k_F^2$ and $u_2 = \frac{1}{24}\rho_0 k_s^4/k_F^2$, and, therefore, we have $g = k_F^2/(24\rho_0)$. For the same reason, the only possible tetrahedral breather mode in three dimensions is the standard tetrahedral breather. In this case, $u(\mathbf{p})$ is a uniform distribution inside a standard tetrahedron with side length $k_F$, which gives $u_0 = \rho_0 k_s^3/k_F^3$ and $u_2 = \frac{1}{40}\rho_0 k_s^5/k_F^3$, and, therefore, we have $g = k_F^2/(40\rho_0^{2/3})$.

## APPENDIX D: DETAILS FOR THE NUMERICAL SIMULATIONS

The numerical results presented in Figs. 2–4 are based on simulations of a (generalized) Gross-Pitaevskii equation, which reads

$$i\partial_t \psi = -\frac{\nabla^2}{2m}\psi + U(\mathbf{r})\psi + \frac{D+2}{2} g|\psi|^{4/D}\psi, \qquad (D1)$$

where $D$ is the spatial dimension. The power $4/D$ in the last term keeps the equation scale invariant, and it naturally reduces to a Gross-Pitaevskii equation in dimension $D = 2$. We always normalize $\psi$ by $\int d^D \mathbf{r}|\psi|^2 = N$ with $N$ the total particle number.

Writing $\psi = \sqrt{n} e^{i\varphi}$ and $\mathbf{v} = \nabla\varphi/m$, Eq. (D1) can be recast into Eqs. (6) and (7) with the pressure term defined by

$$\frac{1}{n}\nabla P = \nabla \left( \frac{(D+2)g}{2m} n^{2/D} - \frac{1}{2m\sqrt{n}}\nabla^2 \sqrt{n} \right). \qquad (D2)$$

We note that this pressure term is exactly the scale-invariant pressure given by Eq. (12), provided that the second quantum pressure term in the bracket can be ignored. In practice, we always choose a large $g$ parameter [$g = \frac{1}{3} \times 10^5$ for Fig. 2, $g = 0.8 \times 10^3$ for Fig. 3 (middle row), $g = 0.4 \times 10^3$ for Fig. 3 (last row), and $g = 3.6 \times 10^3$ for Fig. 4] to reduce the effect from the quantum pressure term. The initial states are taken to be the ground states of Eq. (D1) in the corresponding box traps, which can be efficiently calculated via the imaginary-time evolution of Eq. (D1). The side length $L$ of these boxes are $L = 10 a_{\rm ho}$ for Fig. 2, $L = 25.8 a_{\rm ho}$ for Fig. 3 (middle row), $L = 11.9 a_{\rm ho}$ for Fig. 3 (last row), and $L = 28.2 a_{\rm ho}$ for Fig. 4 with $a_{\rm ho} = \sqrt{(m\omega)^{-1}}$ being the harmonic length. After determining the initial state, Eq. (D1) is then solved numerically via a split-step method on orthogonal grids. We note that the initial state chosen in this way is smooth at the boundary, but the solution obtained from our ideal-gas dynamics has discontinuity at the boundary at $t \to 0$. Thus, the quantum pressure always plays a certain role at the very beginning of the time evolution [20], which causes the visible difference between our numerical results and the ideal-gas solutions shown in Figs. 2 and 3.

## APPENDIX E: VERIFY THE GLE CONDITION FOR EXAMPLE III

With the Wigner function in Eq. (20), we can calculate its first and second moments directly as

$$\langle p \rangle = -n_0 m w (x - ut) c \cos\gamma, \quad (E1)$$

$$\langle p^2 \rangle = n_0 m^2 c^2 \left[ \frac{1}{2} - w(x-ut)\cos^2\gamma \right]. \quad (E2)$$

Then, a straightforward calculation yields

$$\langle p^2 \rangle - \frac{1}{n}\langle p \rangle^2 = mP[n], \quad (E3)$$

with

$$P[n] = \frac{1}{2} g n^2 - \frac{n}{4m} \partial_x^2 \log n, \quad (E4)$$

where $g = mc^2/n_0$. This shows that the Wigner function in Eq. (20) idealizes a hydrodynamic system with the pressure term given by $P[n]$. Moreover, it is straightforward to check that

$$\frac{1}{n}\partial_x P[n] = \partial_x \left( gn - \frac{1}{2m\sqrt{n}} \partial_x^2 \sqrt{n} \right), \quad (E5)$$

where the rhs is the familiar pressure term obtained from the real part of the one-dimensional Gross-Pitaevskii equation (22) by writing $\psi = \sqrt{n} e^{i\varphi}$.